\documentclass{IEEEtran}
\usepackage{cite}
\usepackage{amsmath,amssymb,amsfonts,mathptmx}
\usepackage{algorithmic}
\usepackage{graphicx}
\usepackage{textcomp}
\usepackage{xcolor}
\usepackage{hyperref}
\usepackage[normalem]{ulem}
\usepackage{booktabs}
\usepackage{capt-of}
\usepackage{tabu}

\useunder{\uline}{\ul}{}
\usepackage[linesnumbered,ruled,vlined]{algorithm2e}

\SetCommentSty{mycommfont}

\SetKwInput{KwInput}{Input}                % Set the Input
\SetKwInput{KwOutput}{Output}              % set the Output
\def\BibTeX{{\rm B\kern-.05em{\sc i\kern-.025em b}\kern-.08em
    T\kern-.1667em\lower.7ex\hbox{E}\kern-.125emX}}
\begin{document}
\title{METAHEURISTIC APPROACH TO SOLVE PORTFOLIO SELECTION PROBLEM}
\author{
    \IEEEauthorblockN{Taylan Kabbani\IEEEauthorrefmark{1}\IEEEauthorrefmark{2}}\\
    \IEEEauthorblockA{\IEEEauthorrefmark{1}Huawei Turkey R\&D Center, Istanbul, Turkey
    \\taylan.kabbani2@huawei.com}\\
    \IEEEauthorblockA{\IEEEauthorrefmark{2}Graduate School of Engineering and Science Özyeğin University, Istanbul
    \\taylan.kabbani@ozu.edu.tr}

\thanks{The project's data and code are on \textcolor{blue}{\href{https://github.com/taylankabbani/METAHEURISTIC-APPROACH-TO-SOLVE-PORTFOLIO-SELECTION-PROBLEM}{Github}}}\label{code}}
\maketitle
\begin{abstract}
In this paper, a heuristic method based on Tabu Search and TokenRing Search is being used in order to solve the Portfolio Optimization Problem. The seminal mean-variance model of Markowitz is being considered with the addition of cardinality and quantity constraints to better capture the dynamics of the trading procedure, the model becomes an NP-hard problem that can not be solved using an exact method. The combination of three different neighborhood relations is being explored with Tabu Search. In addition, a new constructive method for the initial solution is proposed. Finally, I show how the proposed techniques perform on public benchmarks.
\end{abstract}
\section{Introduction}
\label{sec:introduction}
Financial markets are at the heart of the modern economy and provide an avenue for selling and purchasing assets such as bonds, stocks, foreign exchange, and derivatives. The prime objective of any investor when investing in any of these markets is to minimize the risk involved in the trading process and maximize the profits generated, this objective can be met by diversifying the amount of money to be invested (Capital) among a set of available assets. Each one of the different ways to choose assets and diversify the capital between them is called a \textit{portfolio}(or \textit{Securities}) and the problem of finding this optimal portfolio that maximizes profits and minimizes risk is called \textit{Portfolio Optimization Problem} (POP) or Portfolio Selection Problem.
In his seminal work, \textit{Markowitz} \cite{1} proposed the Mean-Variance model to solve the portfolio optimization problem, which assumes that the risk of a particular portfolio can be minimized by minimizing the co-variance between the portfolio assets, i.e., minimizing the chance of loss by choosing assets that do not tend to move together (low co-variance), so if one asset's price is going down the other asset will still have a chance to make up for the losses. Despite the pioneer model of \textit{Markowitz}, it does not reflect the real-life trading procedure by lacking cardinality constraints that set a limit to the number of assets allowed in a portfolio as well as lacking any bounding constraints that set limits to the amount of money allowed to be invested in each asset.\\
In this paper, an extension of \textit{Markowitz} \cite{1} model is being considered, in which cardinality constraints and quantity constraints are imposed to make it more adherent to real-world trading mechanisms. The addition of these constraints will turn the model from \textit{Quadratic Programming} (QP) problem to a \textit{Mixed Integer Quadratic Programming} (MIQP) problem, which is an NP-Hard problem that can not be solved using an exact method, therefore a  Metaheuristic approach is needed.\\
The use of local search techniques for the POP has been proposed by \textit{Chang et al.} (2000).\cite{2}. Here, I investigate the possible improvement of their \textit{Tabu Search} (TS) technique by first, proposing an alternative of Initial Solution generation (Constructive heuristics) based on the \textit{Sharp Ration} rather than randomly assigning the initial solution. Second, I introduce an additional move to diversify the search space and escape the local optima also instead of using a fixed step size, different step sizes are being considered, and finally, I investigate the application of the \textit{Token-Ring} method with the \textit{Tabu search algorithm} to implement Multi-Neighborhood Search. The proposed techniques are tested on the benchmarks proposed by \textit{Chang et al.} (2000).\cite{2} which represent real stock markets.

\section{Related Work}
\label{sec:Related Work}
\textbf{Chang et al.} \cite{2} used three different metaheuristic approaches (Genetic Algorithm, Simulated Annealing, and Tabu Search) to solve a multi-objective portfolio optimization problem with cardinality \& quantity constraints. They reported that no individual heuristic was found to be dominating across five data sets. \textbf{Schaerf et al.} \cite{3} explored the use of Tabu Search on single-objective constrained POP, and also proposed new algorithms that combine different neighborhood relations. They implemented several token-ring procedures in a way to diversify and intensify the search space and they concluded that TS is the most promising heuristic to solve POP. In an attempt of making the mean-variance model more realistic, \textbf{Soleimani et al.}\cite{4} introduced sector capitalization as an additional constraint to reduce investment risk, they used GA and compared the results of a small instance problem to the results found by LINGO (which can generate global optima for small problems) and reported that the difference between LINGO’s global optimum and GA’s best objective is just 2.9\%. By combining local search techniques (SD \& FD) and exact mathematical programming (Quadratic programming solver) \textbf{Gaspero et al.} \cite{5} proposed a hybrid local search algorithm, according to their results, the developed solver finds the optimal solution in several instances and is at least comparable to other state-of-the-art methods from the others.\\
Surveying the techniques used in literature (\cite{2}, \cite{3}, \cite{4}, \cite{7}, \cite{8}, \cite{9}) to solve the constrained portfolio optimization problem reveals that Genetic Algorithm is the most used technique to solve the problem, where a couple of studies (\cite{3} ,\cite{4}) considered using Tabu search.

\section{Portfolio Optimization Problem}
\label{sec:Portfolio Optimization Problem}
In this section, the portfolio selection (or optimization) problem is introduced in stages. The unconstrained \textit{Markowitz model} is represented first in Section \ref{3.1}. Subsequently, I introduce the specific constraints of the formulation considered in this study in Section \ref{3.2}. Finally, the Efficient Frontier is explained in Section \ref{3.3}.

\subsection{Unconstrained Markowitz model}
\label{3.1}
Given a set of n available assets, $ A = \{a_{1}, a_{2},...,a_{n}\}$, let $\mu_{i}$ be the expected return  (increase in asset price) for asset $a_{i}$ and $x_{i}$ the portion of the capital invested in asset $a_{i}$, and each pair of assets $(a_{i}, a_{j})$ has a real value of co-variance $\sigma_{ij}$. Let $\mu_{p}$ represent the desired return from the portfolio, thus the formulation of the unconstrained problem is given as:
\begin{equation}
min \sum_{i=1}^n \sum_{j=1}^n x_{i} x_{j} \sigma_{ij}
\end{equation}
\\$$S.t.\,\,\,\,\,\,\,\,\,\,\,\,\,\,\,\,\,\,\,\,\,\,\,\,\,\,\,\,\,\,\,\,\,\,\,\,\,\,\,\,\,\,\,\,\,\,\,\,\,\,\,\,\,\,\,\,\,\,\,\,\,\,\,\,\,\,\,\,\,$$
\begin{equation}
\sum_{i=1}^n \mu_{i} x_{i} \geq \mu_{p}
\end{equation}

\begin{equation}
\sum_{i=1}^n x_{i} = 1 
\end{equation}
\begin{equation}
0 \leq x_{i} \leq 1 \,\,\,\, i = 1,2,...,n
\end{equation}
Eq. (1) minimizes the total variance (risk) associated with the portfolio, Eq. (2) ensures that
the portfolio has an expected return of or bigger than $\mu_{p}$. Eq. (3) ensures that 100\% of the capital is being invested and short-selling is not feasible.

\subsection{Constrained Markowitz model}
\label{3.2}
While the previous single-objective model considers only minimal risk for a given expected return it's desirable to also seek a maximum return for a given expected level of risk, this can be achieved by subsuming the expected return constraint into the objective function via a weighting approach which yields a profitable algorithmic approach. The objective function in Eq.(1) will become Eq.(5), where $\lambda$ represents the risk aversion of the investor, when set to 0 the investor is taking high risk in order to maximize the profits, if 1, the investor is not willing to take any risk but profits will not be necessarily maximized (Eq.(10))\\
The cardinality constraint will determine the number of assets k the portfolio must include. A binary variable $z_{i}$ is introduced to denote whether an asset is being selected or not. Eq.(7) \& Eq.(11). Whereas quantity constraint will set boundaries for the weights (proportions) of included assets by specifying a lower ($\epsilon$) and upper ($\delta$) bounds allowed for the allocated proportions to each asset in the portfolio, Eq.(8).
\begin{equation}
min \,\,\, \lambda [\sum_{i=1}^n \sum_{j=1}^n x_{i} x_{j} \sigma_{ij}] - (1 - \lambda) \sum_{i=1}^n \mu_{i} x_{i}
\end{equation}
\begin{equation}
\sum_{i=1}^n x_{i} = 1 
\end{equation}
\begin{equation}
\sum_{i=1}^n z_{i} = K
\end{equation}
\begin{equation}
\epsilon z_{i} \leq x_{i} \leq \delta z_{i}\,\,\,\,\,\, i = 1,2,...,n
\end{equation}
\begin{equation}
0 \leq x_{i} \leq 1\,\,\,\,\,\ i = 1,2,...,n
\end{equation}
\begin{equation}
0 \leq \lambda \leq 1 
\end{equation}
\begin{equation}
 z_{i} = 
\begin{cases}
    1,&  \text{if asset i is held}\\
    0,              & \text{otherwise} 
\end{cases}
\end{equation}
\begin{equation}
z_{i} \in \{1,0\}, \,\,\,\,\,\ i = 1,2,...,n
\end{equation}
\\Where:
	\begin{itemize}
		\item[*] n: number of stocks in the dataset.
		\item[*] $\lambda$ : The risk aversion of the investor
		\item[*] $x_{i}$ : Proportion of capital invested in asset i
		\item[*] $\sigma_{ij}$ : The co-variance between asset i and j
		\item[*] $\mu_{i}$ : The expected return of the asset i
		\item[*] K: the number of assets in the portfolio 
		\item[*] $\epsilon$ : The minimum proportion invested in asset
		\item[*]$\delta$ : The maximum proportion invested in asset
	\end{itemize}
This formulation is a multi-objective mixed quadratic and integer programming problem for which exact methods do not exist.
\subsection{Efficient Frontier}
\label{3.3}
\begin{figure}[!h]
\centerline{\includegraphics[width=\columnwidth]{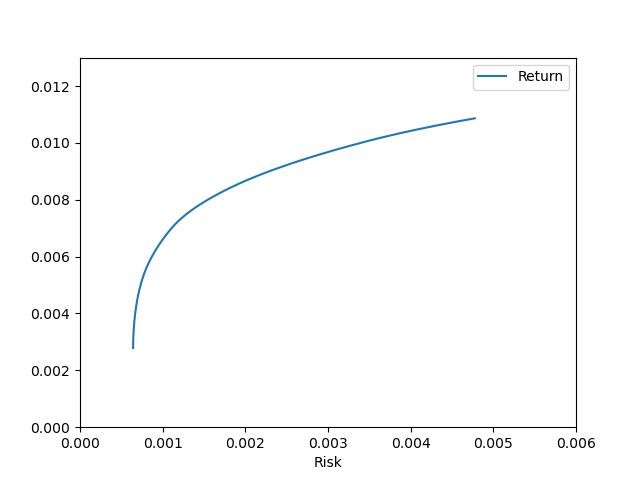}}
\caption{Unconstrained Efficient Frontier corresponding to the smallest benchmark problem. (UEF of Hang Seng)}
\centering
\label{Fig.1}
\end{figure}

Due to the fact that the problem in this study is \textit{multi-objective optimization problem} and usually has several different optimal solutions, it's useful to approach the problem as Pareto Optimality Problem \cite{10}, that is by solving the QP in Section \ref{3.1} for varying values of  $\mu_{p}$ (the desired return from the portfolio) we can trace a non-decreasing curve that represents the set of Pareto-optimal (non-dominated) portfolios, i.e., gives for each expected return the minimum associated risk. We call this curve \textit{The Unconstrained Efficient Frontier} (UEF), Fig.\ref{Fig.1}  shows the UEF for the smallest benchmark problem in \cite{2}, which has been computed taking 2000 different values of $\mu_{p}$ and solving the unconstrained POP. In a similar way, the UEF for each data set in the benchmark is shown in  Table \ref{Table3}. 
Given that the constraint problem has never been solved exactly (Eqs.(5)-(12)), the quality of the heuristic method will be measured in average percentage loss w.r.t. the UEF. i.e., taking the sets of Pareto optimal portfolios obtained by solving the constrained POP using the proposed heuristic approach we can trace out the heuristic efficient frontier and compare it to the UEF (More is Section\ref{5})

\section{Portfolio Optimization by Tabu Search}
\label{4}
For any local search method such as \textit{Tabu Search} (TS), there are essential components we need to define, Initial Solution, Neighborhood Structures, and Search Space. In this section, these components will be described in addition to the TS algorithm parameters used in this study.

\subsection{Search Space and Neighborhood Structures}
In POP, the solution (portfolio) consists of two parts, the first part is the set of assets chosen from the pool of available ones, and the second part is the capital proportions allocated to each chosen asset. Hence, the solution will be represented as two sequences, $L = {[a_{1}, . . . , a_{n}]}$ and $S = {[x_{1}, . . . , x_{n}]}$ such that $a_{i}$ is an asset in the solution (portfolio) and $x_{i}$ is the capital fraction invested in asset $a_{i}$, where the length of L satisfies Eq.(7) and S satisfies Eq.(6), Eq.(8) and Eq.(9).\\
As we are trying to optimize a continuous variable (capital fractions), the neighborhood generation is defined over the \textit{step} of the move performed on the current solution. Given a step size \textit{q} which is a real-value parameter we can define two moves to explore the search space, \textit{Increase\_move} and \textit{Decrease\_move}. The idea is to increase and decrease each of the asset weights in the solution in order to find the optimal ones. To illustrate the aforementioned neighborhood relation further suppose that the current solution is $L = [{2, 8}]$ and $S = {[0.65 , 0.35]}$, the neighborhood solutions following the Increase\_move with step size q = 0,8 would be $S1 = {[(1+0.8) \times 0.65 , 0.35]}$ and $S2 = {[0.65 , (1+0.8) \times 0.35]}$ and the same for the Decrease\_move with replacing (1+ 0.8) by (1- 0.8).\\ \\
\begin{itemize}
\item \textit{Increase\_move (Solution, i, q)}:
	\begin{itemize}
	\item \textbf{Description:} Increases the weight of asset i in Solution. The number of neighbors generated equals to the number of assets k in the solution.
	\item \textbf{Effects:} $x_i = x_i \times (1 + q)$
	\end{itemize}

\item \textit{Decrease\_move (Solution, i, q)}:
		\begin{itemize}
	\item \textbf{Description:} Decreases the weight of asset i in the Solution. The number of neighbors generated equals to the number of assets k in the solution.
	\item \textbf{Effects:} $x_i = x_i \times (1 - q)$, if $ x_i < \epsilon$ replace asset i with a randomly chosen asset j, not in the solution.
	\end{itemize}
\end{itemize}
Notice in the example shown above, the S1 neighbor solution becomes $S1 = {[1.17 ,0.35]}$ which violates the constraints imposed. To ensure that all solutions in the searching space satisfy the imposed constraints (all weights sum up to one and satisfy the upper and lower bounds), \textit{Rescale} (see Algorithm \ref{Rescale}) algorithm is introduced \cite{2}.

In addition to the Increase\_move and Decrease\_move, a third move is added to diversify the search which is the Swap\_move, this move swaps the asset with the lowest weight in the solution by an asset, not in the solution.
\begin{itemize}
\item \textit{Swap\_move (Solution, j)}:
	\begin{itemize}
	\item \textbf{Description:} Swaps asset i which has the lowest weight in Solution by asset j, not in Solution. The number of neighbors generated equals (number of assets available N - number of assets k in the solution)
	\item \textbf{Effects:} Swap(i,j), $ i \in Solution, j \notin Solution$ keep the same weights 
	\end{itemize}
\end{itemize}

%%%% Rescale Algorithm %%%%%%%
\begin{algorithm}[h]
\DontPrintSemicolon
  \KwInput{Solution}
  \KwOutput{Solution which satisfies cardinality and quantity constraints } 
  \tcp{K is \#assets in Solution}
  free\_proportion = $1 - (K \times \epsilon)$ \\
  weight\_sum = $ \sum_{i \in Solution} x_{i}$\\
  \tcp{Fixing weights to add up to 1 and no weight is < $\epsilon$:}
  \For{Each $x_{i}$ in Solution}  
	  		{
	  		$x_{i} = \epsilon + (x_{i}/weight\_sum) \times free\_proportion$
	  		}
\tcp{No weight is > $\delta$:}
R = Assets in Solution with weights  $> \delta$\\
\If {R not empty}
{L = $\sum_{i \in Solution-R} x_{i}$\\
free\_proportion = $1 - (K \times \epsilon + length(R) \times \delta)$ \\
	\For{Each $x_{i}$ in Solution}{ 
		\If {$x_{i} < \delta$}
			{$x_{i} = \epsilon + (x_{i}/L) \times free\_proportion$}
		\Else {$x_{i} = \delta$}
	}}
   \Return {Solution}
	   
\caption{Rescale}
\label{Rescale}
\end{algorithm}
\subsection{Initial Solution Construction}
In the literature \cite{2}, the initial feasible solution is selected as the best among randomly generated portfolios with K assets. In this project, I propose a \textit{Greedy approach} (similar to Fractional Knapsack Problem) based on \textit{Sharpe Ration}\cite{12}, which is basically the ratio of the expected return of an asset to its standard deviation. (See Algorithm \ref{Initial Solution}). To illustrate the effectiveness of the proposed greedy algorithm to find an initial solution, I solve a small problem (toy problem) that had been solved using the exact method, the General Algebraic Modeling System (GAMS) in \cite{13}. As the problem in \cite{13} is an unconstrained Portfolio Optimization Problem (POP), the constraints are relaxed in our model to match with the one solved by \cite{13}, Lambda = 1, Capital = 1000, k=3, epsilon = 0.0, delta = 1, T= 10000, where T is the number of iterations to generate random weights and evaluate them, so by increasing T we might get a better solution but with computation trade-off. The obtained results, i.e., proportions of the capital to be invested in each stock, are very close to the ones found by the exact method, which is a good indication of the efficiency of the proposed approach. Solution found by GAMS : \{0.497 IBM, 0.0 WMT, 0.502 SEHI\}, Solution found by the Greed algorithm : \{0.442 IBM, 0.1036 WMT, 0.454 SEHI\}\textcolor{blue}{(See Code \& Notebook\href{https://github.com/taylankabbani/METAHEURISTIC-APPROACH-TO-SOLVE-PORTFOLIO-SELECTION-PROBLEM/tree/master/InitialSolution}{ here})}
%%%% Algorithm for solution initialization%%%%%%%
\begin{algorithm}[h]
\DontPrintSemicolon
  \KwInput{Expected returns, Standard deviations, Number of iterations to perform(T)}
  \KwOutput{\{stock\_1: $x_{1}$, stock\_2: $x_{2}$,..., stock\_n: $x_{n}$\}}  
  L = [] \tcp*{Starting with empty set of assets}
  S=[]   \tcp*{Starting with empty set of fractions}
  T = Integer \tcp*{Number of specified Iterations}
  Best\_solution = \{\}\\
  V = Infinite \tcp*{Best value found for the objective function}
  \For{Each available asset}  
	  		{
	  		Calculate Sharpe\_ratio = Asset Return / Standard Deviation
	  		}
	  Rank available assets based on their Sharpe\_ratio in a descending order\\
	  Add the first K assets to L  \tcp*{Constraint in Eq.(7)}
  \For {t=1 to T}   
	  { S = \{\} \tcp*{Current solution in the iteration}
	  f = 0    \tcp*{Current value for the objective function}
	  
	  \For{Each asset in L}
	  		{
	  		\tcp{Constraint in Eq.(8) \& Eq.(9)}
	  		$x_{i}$ = randomly assign a value between $\epsilon$ and $\delta$    
	  		}	
	   Renormalize weights in S to add up to one  \tcp*{Constraint in Eq.(6)}
	   \If {$f <V$ \& length(S) = k}
	   {Update the best solution\\
	   V = f\\
	   Best\_solution = S}}
   \Return {L, S}
	   
\caption{Initial Solution}
\label{Initial Solution}
\end{algorithm}
\subsection{Tabu Search Parameters}
The parameters of TS are tuned using the Grid Search technique, i.e., testing different combinations of the parameters and choosing the best-performing one.
\begin{itemize}
\item \textit{Tabu List:} For each neighborhood relation a tabu list is created. Tabu tenure is set to 3 throughout the whole algorithm (static) for Increase\_move and Decrease\_move, whereas for Swap\_move the tabu tenure is 20 due to the fact that this move contributes to the stability of the search.
\item \textit{Tabu Attribute:} Because TS aggressively selects the best admissible solution from the neighborhood space, moves are isolated and updated in short-term memory, which records the increase/decrease of the move value for each solution in the neighborhood. The tabu attribute of Increase\_move and Decrease\_move is the asset  that has been increased/decreased. Swap\_move is the asset that has been added by the move to the solution. For example, [asset5: \{Increase: 0, Decrease: 3, Swap: 20\}] which means that asset5 is tabu to be Decrease for the next three iterations but it's not for Increasing move, and it's tabu to be removed (swapped) if it has the lowest weight in solution for the 20 next iterations.

\item \textit{Aspiration Criteria:} If the current move is tabu but delivers a better solution than the best found one, its tabu statute is overridden.
\item \textit{Termination Condition:} When a solution that improves the Best (the incumbent) solution cannot be obtained in 200 consecutive iterations.
\end{itemize}

\section{Token Ring Search}
The token-ring search is a sequential solving strategy for combining different neighborhood functions. Given an initial state and a set of algorithms or what we call \textit{runners}, the token-ring search makes circularly a run of each runner, always starting from the best solution found by the previous one. The effectiveness of the token-ring search for two runners has been stressed by several authors \cite{11}.

\subsection{t1 Runner}
Given a step size q, the \textit{t1 runner} uses the TS algorithm described in Section \ref{4} to search the space for the local optima solution. That is, given an initial solution $S_{0}$, t1 will iterate to update the current and best solutions, by choosing a \textit{non-tabu move} or \textit{Best-improving move} from the \textit{Neighborhood Union} of the three neighborhood relations applied on the current solution. Algorithm \ref{t1_TS} shows the details of the t1-TS runner.

%%%% t1-TS Algorithm %%%%%%%
\begin{algorithm}[h]
\DontPrintSemicolon
  \KwInput{Initial\_Solution, q}
  \KwOutput{local optima solution found using step size q} 
  Best\_Solution = Initial\_Solution\\
  Current\_Solution = Initial\_Solution\\
  \tcp{Generate neighborhood solutions from the Current\_Solution:}
  Neighbors\_Solutions = [] \\
  \For{asset i in Current\_Solution}
  	{Add Increase\_move (Current\_Solution, i, q) \& 
  	Decrease\_move ((Current\_Solution, i, q) to Neighbors\_Solutions}
 \For {asset j not in Current\_Solution}
 	{Add Swap\_move (Current\_Solution, j) to Neighbors\_Solutions}
Select admissible move from  Neighbors\_Solutions: Select the \textit{non-tabu improving} or \textit{least Improving} move. Override tabu statues if the move meet the \textit{aspiration criteria}.\\
Update Current\_Solution \& Best\_Solution\\
Check Termination:\\
\If {No Best\_Solution found in 200 consecutive iteration} {Return Best\_Solution}
\Else {Go to step 3}
 	
\caption{t1\_TS}
\label{t1_TS}
\end{algorithm}

\subsection{t2 Runner}
In the t2 runner, the step size q is defined. Step Size q starts with a big value (q = 5) and decreases by 0.2 for each run of the t1 runner until it reaches 0.2. By starting with big steps we guarantee \textit{diversification} by making big moves and small steps provide \textit{intensification} with small moves. Here, I define one run of the t2 runner as a complete  iteration over all step sizes using the t1 runner. t2 runner stops when no improving solution is found by the current run (See Algorithm \ref{t2}).  t1\_TS always starts from the best solution found by the previous one.

%%%% t1-TS Algorithm %%%%%%%
\begin{algorithm}[h]
\DontPrintSemicolon
  \KwInput{Initial\_Solution}
  \KwOutput{Best Solution} 
  Initialize: step\_size\_list = $[q_{1}=5,q_{2}=4.8,...,q_{n}=0.2 ]$\\
  Initial\_Solution = Initial Solution Algorithm \\
  Next\_Solution=t1\_TS(Initial\_Solution, $q_{0}=5.2$) \tcp*{\ref{t1_TS}}
  \For {step $q_{i}$ in step\_size\_list \tcp*{One r2 run}} 
  	{Reset tabu time of all moves\\
  	Next\_Solution = t1\_TS (Next\_Solution , $q_{i}$)}
  \If {Improved solution found by any step\_size in 4}{Go to 4 \tcp*{another t2 run}}
  
  \Else {\tcp{Final Soluiton}Stop \& return Next\_Solution}
  	
\caption{t2}
\label{t2}
\end{algorithm}

\section{Computational Experiments}
\label{5}
\subsection{Benchmarks and experimental setting}
\begin{figure}[!h]
\centerline{\includegraphics[width=\columnwidth]{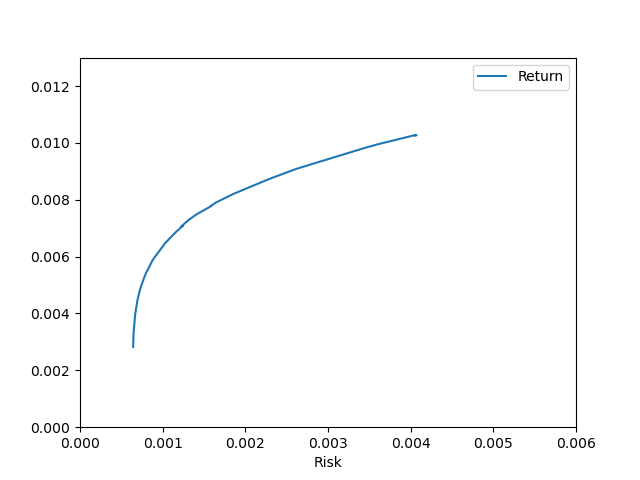}}
\caption{Constrained Efficient Frontier corresponding to the smallest benchmark problem.  (UEF of Hang Seng)}
\centering
\label{Fig.2}
\end{figure}

In this section, the proposed approach in this study is being used to solve the Constrained Model described in Section\ref{3.2}. I have done all the computational experiments with five sets of benchmark data that have been already used in \cite{2}. These data correspond to weekly prices from March 1992 to September 1997 and they come from the indices: Hang Seng in Hong Kong, DAX\_100 in Germany, FTSE\_100 in the UK, S\&P\_100 in the USA, and Nikke\_225 in Japan. The number of assets varies within each data set: 31, 85, 89, 98, and 225 assets, respectively. \href{ http://people.brunel.ac.uk/~mastjjb/jeb/orlib/portinfo.html)}{Data Source}\\
All the results presented here have been computed as in \cite{2}, using the values $ K = 10, \epsilon = 0.01
and \delta = 1 $ for the problem formulation, and the values $\Delta \lambda = 0.02 $ for the
implementation. So the method is implemented with 51 different values for the risk aversion parameter $\lambda$. As a result for each benchmark data we have 51 optimal solutions (Portfolios), and each has the best objective function value found by the proposed heuristic for the weighting parameter $\lambda$. From these obtained portfolios (solutions) we can trace back the \textit{Constrained Efficient Frontier} (CEF) by calculating the \textit{Return} and \textit{Variance of return} (Risk) of each portfolio. Fig \ref{Fig.2} shows the CEF for the smallest benchmark data, where 51 points are plotted, and each point represents a solution (Portfolio) with (x=Risk,y = Return). Similarly, Table \ref{Table3} shows the UEF for each data set in the benchmark.
\begin{table*}[ht]
\centering
\caption{Comparison of the proposed method performance with benchmark TS algorithm}
\label{Table}
\resizebox{\textwidth}{!}{%
\begin{tabular}{|ccccc|}
\hline
\multicolumn{1}{|c|}{Index} &
  \multicolumn{1}{c|}{Assets} &
  \multicolumn{1}{c|}{} &
  \multicolumn{1}{c|}{TS\&TokenRing} &
  TS In Ref{[}2{]} \\ \hline
Hang Seng &
  31 &
  \begin{tabular}[c]{@{}c@{}}Median percentage error\\ Mean percentage error\\ Time (s)\end{tabular} &
  \begin{tabular}[c]{@{}c@{}}1.812\\ 2.2656\\ 1154.3\end{tabular} &
  \begin{tabular}[c]{@{}c@{}}1.2181\\ 1.1217\\ 74\end{tabular} \\ \hline
DAX &
  85 &
  \begin{tabular}[c]{@{}c@{}}Median percentage error \\ Mean percentage error \\ Time (s)\end{tabular} &
  \begin{tabular}[c]{@{}c@{}}4.21\\ 4.035\\ 2873\end{tabular} &
  \begin{tabular}[c]{@{}c@{}}2.6380\\ 3.3049\\ 199\end{tabular} \\ \hline
FTSE &
  89 &
  \begin{tabular}[c]{@{}c@{}}Median percentage error \\ Mean percentage error \\ Time (s)\end{tabular} &
  \begin{tabular}[c]{@{}c@{}}1.2406\\ 1.2959\\ 2919\end{tabular} &
  \begin{tabular}[c]{@{}c@{}}1.0841\\ 1.6080\\ 246\end{tabular} \\ \hline
S\&P &
  98 &
  \begin{tabular}[c]{@{}c@{}}Median percentage error \\ Mean percentage error \\ Time (s)\end{tabular} &
  \begin{tabular}[c]{@{}c@{}}2.3630\\ 2.5068\\ 3107\end{tabular} &
  \begin{tabular}[c]{@{}c@{}}1.2882\\ 3.3092\\ 225\end{tabular} \\ \hline
Nikkei &
  225 &
  \begin{tabular}[c]{@{}c@{}}Median percentage error \\ Mean percentage error \\ Time (s)\end{tabular} &
  \begin{tabular}[c]{@{}c@{}}1.34635\\ 1.21220\\ 5866.2\end{tabular} &
  \begin{tabular}[c]{@{}c@{}}0.6093\\ 0.8975\\ 545\end{tabular} \\ \hline
\end{tabular}%
}
\end{table*}
\subsection{Percentage Deviation Calculation}
To measure and quantify the quality of the solutions obtained, hence the performance of the proposed method,  the percentage deviation (distance) of each portfolio from the (linearly interpolated) \textit{Unconstrained Efficient Frontier} (UEF) is computed as following:\\
Let the pair $(v_{i}, r_{i})$ represent the variance and mean the return of a point in the CEF (solution found by the proposed method). Let also $(\hat{v_{i}},\hat{r_{i}})$ represent the variance and mean the return of a point in the UEF.
We can find $v_{j}$ and $v_{k}$ which brackets $v_{i}$ on the UEF as, $v_{j} = \min[\hat{v_{i}} | \hat{v_{i}} \geq v_{i}] $ and $v_{k} = \max[\hat{v_{i}}| \hat{v_{i}} \leq v_{i}] $ and by simple geometry we can calculate $\hat{\hat{r_{i}}}$  which is the value associated with the x-direction linear interpolated point with $\hat{v_{i}} = v_{i}$ on the UEF.  $\hat{\hat{r_{i}}} =  r_{k} + (r_{j} - r_{k})[(r_{i} - r_{k})/(r_{j} - r_{k})] $ . 
From $\hat{\hat{r_{i}}}$ we can calculate convenient percentage deviation error of the solution as $\varphi_{i} = | 100 \times (r_{i} - \hat{\hat{r_{i}}})/ \hat{\hat{r_{i}}}|$. The percentage deviation error associated with y-direction is calculated in the same way but with keeping the Return $\hat{r_{i}}$ fixed instead of $\hat{v_{i}}$. The percentage deviation error measure of solution $(v_{i}, r_{i})$ is the minimum of the x-direction, and y-direction percentage deviations. Finally the Median percentage error and Mean percentage error of the 51 solutions found by the proposed method are reported and compared with reference \cite{2} for each data set.

\subsection{Results \& Conclusion}
The results for the proposed heuristic algorithm (Tabu Search with Token-Ring technique) with the number of assets in the portfolio K = 10, a minimum proportion of any asset that must be held (if any is held) $\epsilon = 0.01$ and maximum proportion that can be held of any asset (if any is held) $\delta = 1$ are shown in Table \ref{Table}. \\
The results show that the proposed method's performance is very close to the benchmark approach. For the problems with assets = 98 and 89, the proposed method surpasses the benchmark results in the Mean percentage error, whereas for the rest of the data sets, it delivers sub-optimal results in comparison to the benchmark.\\
The major drawback of the proposed method in this study is the computation time which was significantly higher than the benchmark as shown in Table \ref{Table}.  In conclusion, the proposed method of combining the Token Ring technique with three neighborhood relations failed to surpass the benchmark for some data instances and successfully outperformed it for other data instances. To better analyze the performance of the proposed method, the average error in the first objective (Variance return) and the average error in the second objective (Return) is being reported in Table \ref{Table2}. Here, The variance of return (Risk) error for any heuristic point $(v_{i}, r_{i})$ is computed as the value $100 \times (\hat{v_{i}} - v_{i}) / \hat{v_{i}}$, where $\hat{v_{i}}$ is the corresponding to Risk with keeping Return $r_{i}$ fixed according to the linear interpolation in the UEF. The average value of all the errors for the points in heuristic CEF gives us the variance of return error shown in Table \ref{Table2}. Similarly, the average Return error is calculated. The analysis indicates that the studied method is performing well for the second objective (Return) and relatively poor with the first objective (Risk), that is for small values of lambda (giving more weight to the second objective) the method performs well, i.e., the method is performing well if we are considering portfolios with high return and small risk.

\begin{table}[ht]
\centering
\caption{Comparison between the errors of the first objective and second objectoive }
\label{Table2}
\resizebox{\columnwidth}{!}{%
\begin{tabular}{|cccc|}
\hline
\multicolumn{1}{|c|}{Index} &
  \multicolumn{1}{c|}{Assets} &
  \multicolumn{1}{c|}{} &
  TS\&TokenRing \\ \hline
Hang Seng & 31  & \begin{tabular}[c]{@{}c@{}}First Obj (Risk) error\\ Second Obj (Return) error\end{tabular}  & \begin{tabular}[c]{@{}c@{}}7.0405\\ 2.4751\end{tabular} \\ \hline
DAX &
  85 &
  \begin{tabular}[c]{@{}c@{}}First Obj (Risk) error \\ Second Obj (Return) error\end{tabular} &
  \begin{tabular}[c]{@{}c@{}}49.7685\\ 1.4008\end{tabular} \\ \hline
FTSE &
  89 &
  \begin{tabular}[c]{@{}c@{}}First Obj (Risk) error \\ Second Obj (Return) error\end{tabular} &
  \begin{tabular}[c]{@{}c@{}}6.0545\\ 1.8224\end{tabular} \\ \hline
S\&P &
  98 &
  \begin{tabular}[c]{@{}c@{}}First Obj (Risk) error \\ Second Obj (Return) error\end{tabular} &
  \begin{tabular}[c]{@{}c@{}}20.2107\\ 3.5328\end{tabular} \\ \hline
Nikkei    & 225 & \begin{tabular}[c]{@{}c@{}}First Obj (Risk) error \\ Second Obj (Return) error\end{tabular} & \begin{tabular}[c]{@{}c@{}}9.221\\ 2.2761\end{tabular}  \\ \hline
\end{tabular}%
}
\end{table}

\clearpage
\begin{table}[ht]
\caption{The Unconstrained Efficient Frontier (UEF) and the Constrained Efficient Frontier (CEF) of each data set. The Return (second obj) is represented on the y-axis, whereas the Risk (first obj) is represented on the x-axis}
\label{Table3}
\centering
\begin{tabu}to \columnwidth {X[c]X[c]}
  \includegraphics[width=45mm]{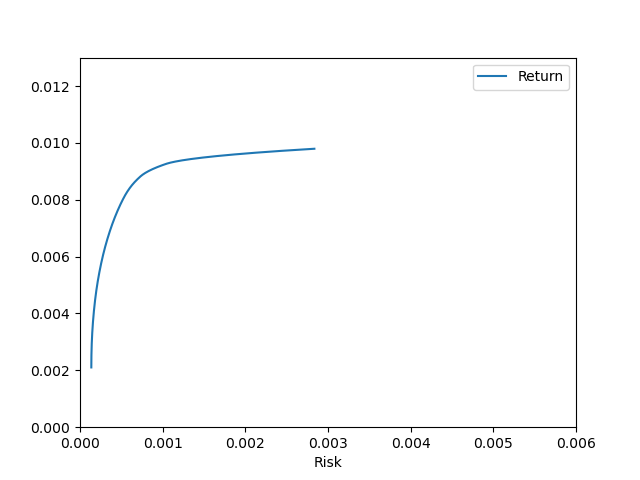}\captionof{figure}{UEF of DAX} &\includegraphics[width=45mm]{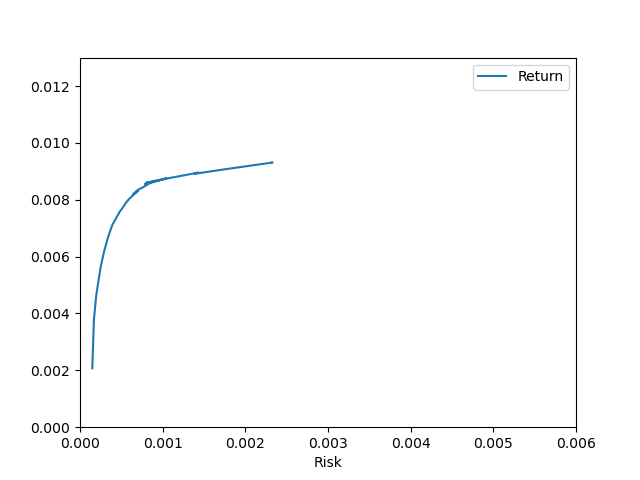}\captionof{figure}{CEF of DAX} \\
  \includegraphics[width=45mm]{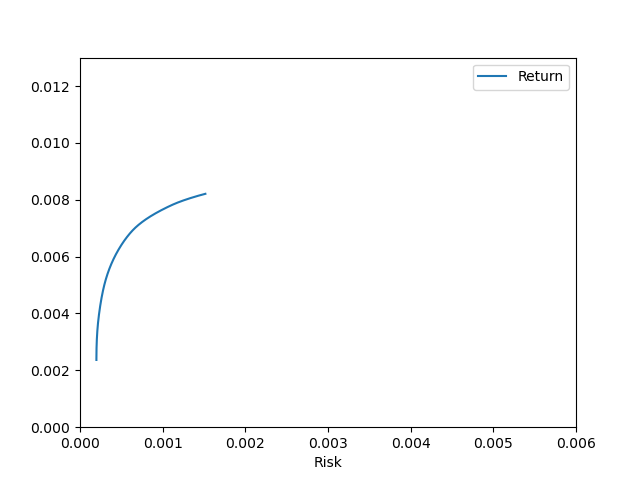}\captionof{figure}{UEF of FTSE} &\includegraphics[width=45mm]{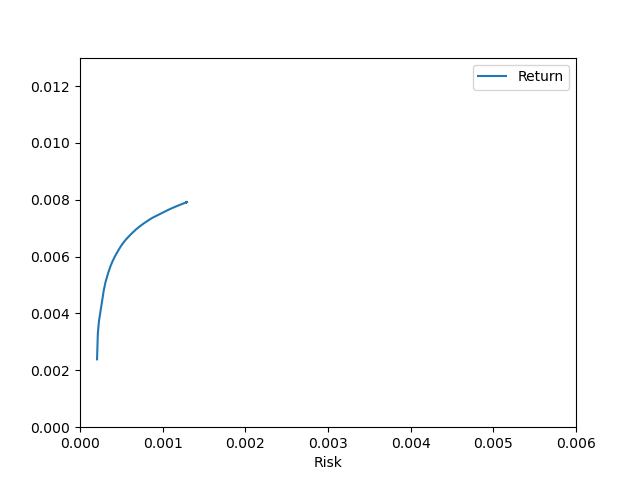}\captionof{figure}{CEF of FTSE} \\
  \includegraphics[width=45mm]{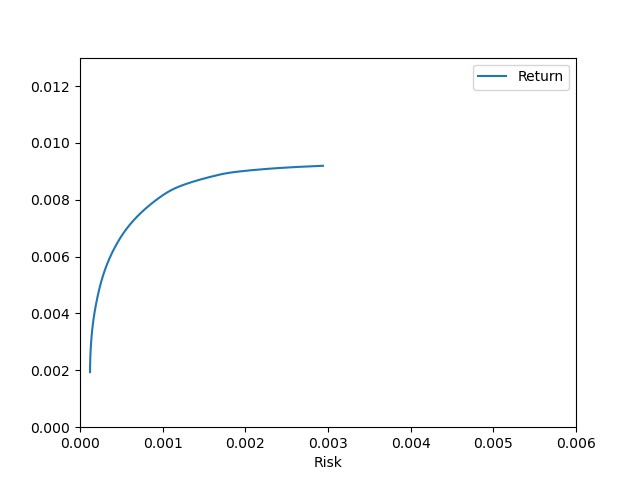}\captionof{figure}{UEF of S\&P} &\includegraphics[width=45mm]{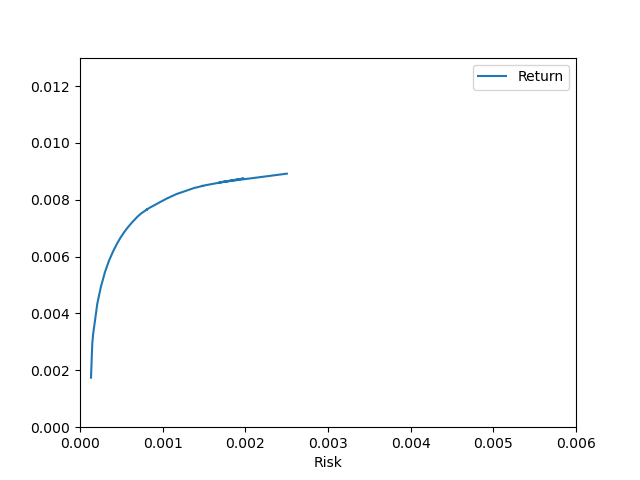}\captionof{figure}{CEF of S\&P} \\
  \includegraphics[width=45mm]{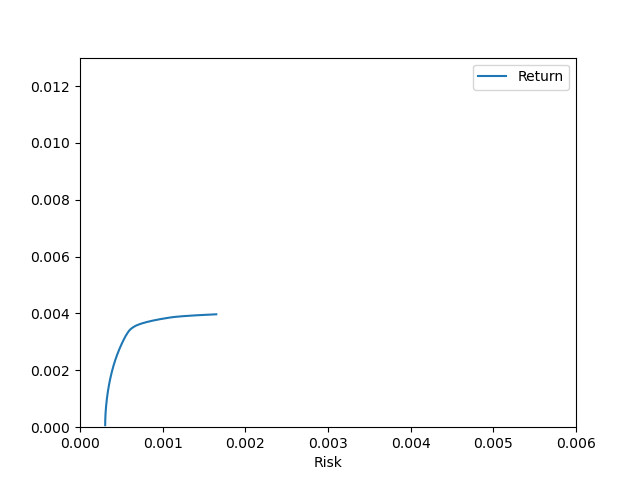}\captionof{figure}{UEF of Nikkei} &\includegraphics[width=45mm]{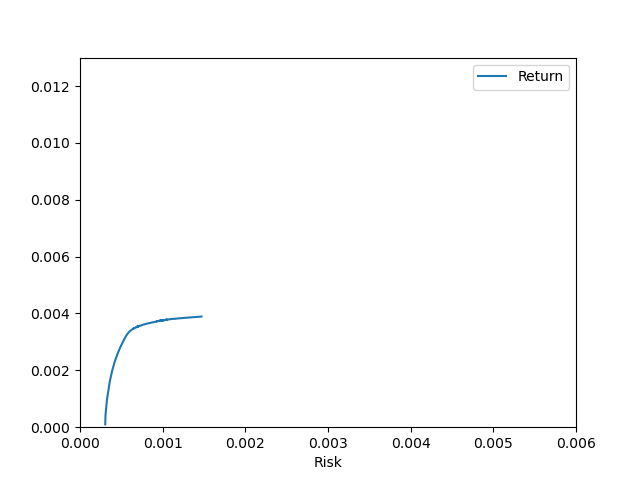}\captionof{figure}{CEF of Nikkei} \\
\end{tabu}
\end{table}

\end{document}